\documentclass{aa} 
\usepackage{graphicx}
\begin{document}
   \title{Extragalactic neutrino background from very young pulsars
surrounded by supernova envelopes}

   \author{W. Bednarek}

   \offprints{W. Bednarek}

   \institute{Department of Experimental Physics, University of \L \'od\'z,
              90-236 \L \'od\'z, ul. Pomorska 149/153, Poland\\
             \email{bednar@fizwe4.fic.uni.lodz.pl}
             }

   \date{Received ...., .... ; accepted ...., ....}

   \abstract{
We estimate the extragalactic muon neutrino background which is produced
by hadrons injected by very young pulsars at an early
phase after supernova explosion. It is assumed that hadrons are accelerated 
in the pulsar wind zone which is filled with thermal photons captured below
the expanding supernova envelope. In collisions with those thermal photons
hadrons produce pions which decay into muon neutrinos.
At a later time, muon neutrinos are also produced by the hadrons in
collisions with matter of the expanding envelope. We show that extragalactic
neutrino background predicted by such a model should be detectable
by the planned 1 km$^2$ neutrino detector if a significant part of pulsars
is born with periods shorter than $\sim 10$ ms. Since such population of
pulsars is postulated by the recent models of production of
extremely high energy cosmic rays, detection of neutrinos with predicted
fluxes can be used as their observational test.    
   }

\titlerunning{Extragalactic neutrino background from pulsars}

   \maketitle
%

\section{Introduction}

The construction of high energy neutrino telescopes with the effective area 
$>10^5$ m$^2$ (AMANDA$\rightarrow$IceCube, ANTARES, NESTOR) 
will probably allow observation of first neutrinos of astrophysical origin,
giving for the first time clear information about high energy hadronic
processes in the universe. Therefore estimations of neutrino fluxes from
discrete sources and  their contribution to the extragalactic neutrino
background (which may be probably easier to detect) is so important.  

Up to now the flux of extragalactic neutrino background has been estimated
with rather high confidence in the case of interaction of diffuse cosmic
rays with the matter and radiation. Estimations of neutrino fluxes from
discrete sources, where as it is well known, high energy processes 
are important i.e. active galactic nuclei and $\gamma$-ray bursts, are 
less certain. Also more
exotic processes, like decays of topological defects, have been discussed as
possible sources of extremely high energy neutrinos.  For recent reviews of
these estimations we refer to the articles by Protheroe~(1998)
and Learned \& Mannheim~(2000).

Supernova explosions, during which young pulsars are formed, seem to be
another likely sites of high energy processes in which observable neutrino
fluxes are expected (e.g. Berezinsky \& Prilutsky~1978; Bednarek \&
Protheroe~1997; Protheroe, Bednarek \& Luo~1998). In our recent paper (Beall \&
Bednarek~2001, BB) we calculate the muon neutrino (and antineutrino) flux
expected from a very young pulsar surrounded by the supernova envelope 
in our Galaxy. It has been shown that muon
neutrinos are likely to be observed in the large scale (1 km$^2$) neutrino
detector from the supernova within our
Galaxy if a pulsar with a period shorter than $\sim 10$ ms is formed. 
However such explosions happen relatively rarely and we would have to be
lucky to detect such a supernova within reasonable period of time. Therefore in
order to test the predictions of the mentioned model by the planned neutrino
detectors, we calculate in this paper the extragalactic
neutrino flux produced by very young pulsars formed in the universe.

\section{Neutrinos from nuclei accelerated by pulsars}

Pulsars are produced during the type II and Ib/c supernova explosions of
massive stars which detailed models have been investigated in many papers (see
e.g. Woosley et al.~1993). When the iron core collapses, a very hot
protoneutron star is formed. It cools to a neutron star within about $t_{\rm
NS}\approx 5-10$ s from the collapse (Burrows \& Lattimer~1986, Wheeler et
al.~2000). The rest of the presupernova mass is expelled
with the velocity of the inner radius of the order of $v_1 = 3\times 10^8$ 
cm s$^{-1}$ producing the supernova envelope. Its initial very high
optical depth drops fast with time. The region between the pulsar and the
envelope is filled with thermal radiation of characteristic temperature 
$T_{0}\approx 3\times 10^9$ K (see Fig.~8 in Woosley et
al.~1993). This radiation is not able to escape through the envelope because
of its high optical depth. Assuming that the total energy of radiation inside
the envelope is conserved, we can estimate the temperature of this radiation
as a function of time during the expansion of the envelope    
$T(t) = T_{0} [R_1/(R_1 + v_1(t_{\rm NS} +
t))]^{3/4}$, where $R_1 = 3\times 10^8$ cm is the initial radius of
the envelope at the moment of formation of the protoneutron star.
When the pulsar forms (about 10 s after the explosion), the
radiation temperature inside the envelope already drops to
$\sim 5\times 10^8$ K. This radiation escapes through the envelope when the
optical depth drops below $\sim 10^3$. This happens about $\sim 2\times
10^6$ s after the explosion (see BB).

We assume that at the very early stage the pulsar loses energy only on
electromagnetic waves. Therefore its period changes according to the formula
$P_{\rm ms}^2(t) = 1.04\times 10^{-9} tB_{12}^2 + P_{\rm 0,ms}^2$, where 
$P_{\rm 0,ms}$ $P_{\rm ms}$ are the initial and present periods of the pulsar,
and $B_{12}$ is the pulsar surface magnetic field in units $10^{12}$ G.  
We neglect the pulsar energy losses to gravitational radiation due to the
r-mode instabilities since they may become important about 1 year after
pulsar formation (e.g. Andersson 1998, Lindblom, Owen, Morsink 1998).

It is argued that pulsars can accelerate particles to the highest energies
observed in cosmic rays in the slingshot mechanism (see recent papers by e.g.
Blasi, Epstein \& Olinto~2000 (BEO), De Goubeia Dal Pino \& Lazarian~2000, 
Bednarek \& Protheroe~2001). Following work by Blasi et al.~(BEO),
we assume that a part of the magnetic energy in the pulsar's wind zone
accelerates iron nuclei.  The maximum energies, which nuclei can reach in the
wind zone, depend on the pulsar parameters  
\begin{eqnarray}
E_{\rm Fe} = {{B^2(r_{\rm LC})}\over{8\pi n_{\rm GJ}(r_{\rm LC})}}
\approx 1.8\times 10^{11} \beta B_{12} P_{\rm ms}^{-2}~{\rm GeV}, 
\label{eq2}
\end{eqnarray}
\noindent
where $n_{\rm GJ} = B(r_{\rm LC})/(2 e c P)\approx 3.3\times 10^{11} B_{12}
P_{\rm ms}^{-4}$ is Goldreich \& Julian (1969) density, $r_{\rm LC}\approx
4.77\times 10^6 P_{\rm ms}$ cm,  $c$ is the velocity of light, and 
$\beta$ is the coefficient describing the efficiency of acceleration.
According to the slingshot mechanism the acceleration of nuclei occurs very
fast so that they do not lose energy during this stage. 
After acceleration, the nuclei move balistically through the radiation field 
(below the envelope) and later through the matter of the envelope.
These particles are not frozen into the expanding plasma of the envelope and
therefore do not lose energy on adiabatic process. 

We apply the spectrum of iron nuclei accelerated in the pulsar wind zone close
to the pulsar light cylinder derived in Beall \& Bednarek~(BB). 
In this simple model the number of nuclei
accelerated to energies $E$ scales as a part $\eta$ of the
Goldreich \& Julian (1969) density at the light cylinder radius. 
It is assumed that the pulsar with specific parameters injects particles
within some range of energies which is due to the fact that the magnetic field
at different parts of the light cylinder radius is different. 
The energy spectrum of the iron nuclei below $E < E_{\rm Fe}$, which 
injected by the pulsar at a fixed age 't' of the pulsar, is 
(see details in BB) 
\begin{equation}
\begin{array}{l}
{{dN}\over{dEdt}} = {{2\pi c \eta r_{\rm LC}^2 n_{\rm GJ}
(E_{\rm Fe} E^2)^{-1/3}}\over{3\left[(E_{\rm Fe}/E)^{2/3} -
1\right]^{1/2}}} \\
~~~~~~\cong  {{3\times 10^{30}\eta (B_{12} P_{\rm
ms}^{-2}E^{-1})^{2/3}} \over{\left[(E_{\rm Fe}/E)^{2/3} - 1\right]^{1/2}}}
~{\rm {{Fe}\over{s~GeV}}}.     
\label{eq4}
\end{array}
\end{equation} 

Accelerated nuclei move in the pulsar wind almost at rest
in the wind reference frame (BEO).
Therefore we can neglect their synchrotron energy losses. However, as we
mentioned above nuclei interact with strong thermal radiation field
suffering complete photodesintegration before losing significant energies on
$e^\pm$ pair and pion creation. For the considered parameters of the pulsar and
the acceleration region, nucleons from desintegration of nuclei lose
their energy mainly on pion production. Pions decay into neutrinos if
their decay distance $\lambda_\pi \approx 780\gamma_\pi~~{\rm cm}$, is
shorter than their characteristic energy loss mean free path. The Lorentz
factors of pions, $\gamma_\pi$, are  comparable to the Lorentz factors of
their parent nucleons, so they move similarly in the pulsar wind and
their synchrotron losses do not dominate their inverse Compton scattering
losses (ICS) in the thermal radiation. Pions lose energy on ICS process
mainly in the Klein-Nishina (KN) regime, but not very far from 
the Thomson regime.  By comparing the pion mean free path for ICS losses with
their decay distance  scale $\lambda_\pi$, we have found that pions can decay
before losing energy on ICS if the temperature of thermal radiation drops below
$T\sim 3\times 10^6$ K.  The temperature of radiation below the envelope drops
to this value about $t_{\rm dec}\sim 10^4$ s after supernova explosion. 
At this time nuclei are injected with the maximum energies allowed by 
Eq.~\ref{eq2} and
cool in collisions with thermal radiation mainly on pion production. However
when the optical depth through the expanding envelope drops below $\sim 10^3$
then the radiation is no more confined to the region below the envelope
and its temperature drops much faster. Therefore we conclude that nucleons are
able to cool efficiently in the radiation and produce pions which decay into
muon neutrinos, only from $t_{\rm dec}\approx 10^4$ s up to $t_{\rm
conf}\approx 2\times 10^6$ s after explosion. At later times, the relativistic
nuclei interact with the matter of the envelope which density is already low
enough for the pions to decay into high energy neutrinos.
Relativistic nucleons find enough column density of matter in the envelope ($>
10$ g cm$^{-2}$) to interact at the time shorter than about 1 year after
explosion (see BB). 

In terms of such a model we compute the spectrum of muon neutrinos
from the pion decay, $N(E_\nu, P, B)$, produced by protons accelerated
by pulsars with different periods, $P$, and surface magnetic fields, $B$
(see BB). The pions are produced during the period $10^4 -
2\times 10^6$ s after explosion as a result of complete cooling of nucleons in
collisions with thermal radiation below the envelope, and during the period
$2\times 10^6 - 3\times 10^7$ s from explosion as a result of collisions of
iron nuclei with the matter of the envelope.

\section{Extragalactic neutrino background}

In order to calculate the extragalactic neutrino background (ENB) we assume
that the pulsars born at earier epoch do not differ significantly
from those born at present. Thus the injection spectrum of neutrinos by a
single pulsar depends only on its initial parameters (period, surface magnetic
field) but not on redshift. Under this assumption, we can evaluate the
neutrino injection rate at different redshifts by multiplying the injection
spectrum of neutrinos of an average pulsar in our Galaxy by the neutron
star formation rate, which depends on the massive star formation rate, which
in turn depends on the redshift. In the previous section we have
calculated the injection spectrum of neutrinos from the pulsar
with known parameters (for details see BB). Let's  determine now the
neutrino injection rate from an 'average' pulsar by integrating over the 
initial period and surface magnetic field distributions of pulsars within the
Galaxy, according to 
\begin{eqnarray} 
Q(E_\nu) = \int_{\rm P_{\rm min}}^{\rm P_{\rm max}} 
\int_{\rm B_{\rm min}}^{\rm B_{\rm max}}  N(E_\nu, P, B) f(P) {{dN}\over{dB}}
dP dB, 
\label{eq5}
\end{eqnarray}
\noindent
where $P_{\rm min}\div P_{\rm max}$ is the range of initial periods of 
pulsars, $B_{\rm
min}$ and $B_{\rm max}$ correspond to the range of typical values of the
surface magnetic fields for the observed "classical" radio pulsars equal to
$10^{11.5}$ and $10^{13.3}$, respectively.  

It is believed that pulsars are born with rather short periods. However
their exact period distribution is not known. Therefore we consider two
possible distributions (see Giller \& Lipski~\cite{gl01}). The simplest case
is described by the uniform distribution,  
$f(P) = 1/(P_{\rm max}-P_{\rm min})$.
And probably a more realistic case, 
described above 1 ms by the gamma function, which is defined by the mean
pulsar period $\langle P\rangle$ and the index $\alpha$,
$f(P) = A P^{\alpha} e^{-(\alpha + 1)P/\langle P\rangle}$,
where $A = ((\alpha + 1)/\langle P\rangle)^{(\alpha + 1)}/\alpha!$ after
normalization to one.    

The distribution of surface magnetic fields of pulsars at birth can be 
estimated from the observed population of radio pulsars and assuming that
the pulsar magnetic fields do not decay during the radio activity stage.
It can be approximated by 
$dN/d(\log B)\approx 0.065 B_{12}^{-1}~{\rm year^{-1}}$,
for $B_{12}>2$, after normalization to the formation rate of pulsars in the
Galaxy equal to 1 per 70 years (Narayan~\cite{n87}). We estimate the
distribution of pulsars  with surface magnetic fields below $2\times 10^{12}$
G from Fig.~13 in Narayan~(\cite{n87}). This approximation may
underestimate the number of pulsars with low magnetic fields, because of
observational selection effects, and with high magnetic fields $> 10^{13}$ G
since it does not take into account the so called 'magnetars'.

Now we can calculate the differential flux of ENB 
using the formula (see Fichtel \& Trombka~\cite{ft81}),
\begin{eqnarray}
N_\nu = {{c N_{\rm G}}\over{4\pi H_{\rm o}}}\int_{\rm 0}^{z_{\rm max}}
{{Q(E_\nu(1 + z))F(z)}\over{(1 + z) (1 + q_{\rm 0}z)^{1/2}}}~dz,  
\label{eq9}
\end{eqnarray}
\noindent
where $z$ is the redshift, $N_{\rm G} = 10^{-2}$ Mpc$^{-3}$ is the density of
galaxies,  $Q(E_\nu(1 + z))$ is the injection spectrum of neutrinos with
energy $E_\nu(1 + z)$, averaged over the distribution of initial pulsar
periods and surface magnetic fields (see Eq.~\ref{eq5}). This is normalized to
the pulsar birth rate per galaxy at the redshift $z = 0$ equal to the one
observed in our Galaxy. $f(z)$ is the evolution factor of the pulsar birth
rate  which we take proportional to the supernova rate (Kobayashi, Tsujimoto \&
Nomoto~\cite{ktn00}). Moreover we assume $H_{\rm o} = 75$ km s$^{-1}$
Mpc$^{-1}$, $q_{\rm 0} = 0$, and  $z_{\rm max} = 5$.

The results of calculations of ENB are shown in Figs.~1 and 2, for the case of
initial pulsar periods described by the uniform and gamma distributions,
respectively. We investigate how the ENB depends on the parameters which
define those distributions. In the case of uniform distribution of the initial
periods we show the ENB from pulsars born within the range of periods
limited by $P_{\rm min} = 1$ ms and $P_{\rm max} = 100$ ms,
$P_{\rm min} = 1$ ms and $P_{\rm max} = 10$ ms, and
$P_{\rm min} = 10$ ms and $P_{\rm max} = 100$ ms.
In the case of gamma distribution we show 
the ENB for $\langle p\rangle = 10$ ms and the index $\alpha = 2$, 10 ms
and $\alpha = 4$, and 50 ms and $\alpha = 2$. In all cases $\eta = 1$.
Note that the ENBs, estimated for the pulsars which part has been born with
periods below 10 ms, are above the atmospheric neutrino background
(ANB) produced in the interactions of cosmic ray particles with the earth's
atmosphere. Therefore they can be potentially observed by
future detectors.

\section{Discussion and Conclusion}

We considered only neutrinos produced during the time when the
supernova envelope is optically thick. However pulsars inject particles
also after $> 1$ year when the envelope becomes optically thin. 
Without detailed modeling of propagation of these particles in the
supernova nebula it is difficult to conclude if the neutrino 
fluxes should be limited by the optically thick or thin neutrino bound marked
in Figs.~1 and 2 (Waxman \& Bahcall~1999, Mannheim, Protheore \&
Rachen~2001). The particles injected after 1 year can be captured by the
magnetic field of the expanding large scale supernova nebula and suffer
significant adiabatic energy losses (e.g. Bednarek \& Protheore~2001). The
arguments that charged particles convert into neutrons, e.g. protons into
neutrons, (see Waxman \& Bahcall~1999) may not be valid if the optical depth
for hadronic collisions becomes low and adiabatic losses of captured hadrons
are large.

%
%
   \begin{figure}
   \centering
   \includegraphics[angle=0,width=10.cm]{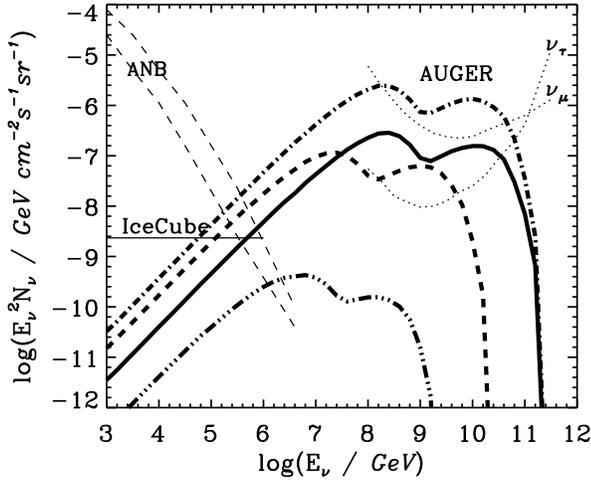}
      \caption{The extragalactic neutrino background produced by 
pulsars with uniform distribution of initial periods
between 1-10 ms (dot-dashed curve), 1-100 ms (full
curve), 10-100 ms (dot-dot-dot-dashed curve) and the acceleration
parameter $\beta =1$, and 1-10 ms and $\beta = 0.1$ (thick dashed curve).
The upper bounds on extragalactic
neutrinos for the case of optically thick (MP\&R, Mannheim et al.~2001) and
thin sources (W\&B, Waxman \& Bahcall~1999)
are marked by the horizontal thin dot-dashed lines. The atmospheric
neutrino background (ANB) is shown by the dashed curves. The sensitivity of
the planned IceCube detector is marked by the horizontal thin full line
(Spiering~\cite{s00}), and the AUGER array by the thin dotted curves for the
muon $\nu_\mu$ and tau $\nu_\tau$ neutrinos (Bertou et al.~\cite{betal01}).}
    \label{fig1}
    \end{figure}
%
%

%
   \begin{figure}
   \centering
   \includegraphics[angle=0,width=10.cm]{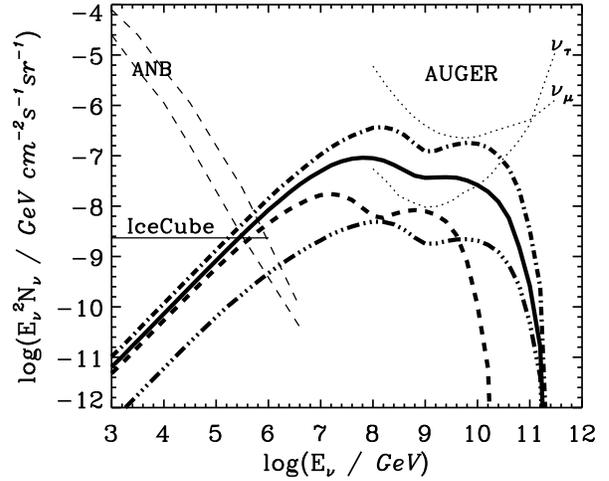}
      \caption{As in Fig.~1 but for pulsars with the initial periods 
described above 1 ms by the gamma function with the mean period 
$\langle p\rangle = 10$ ms
and index $\alpha = 2$ (dot-dashed curves), 10 ms and $\alpha = 4$ (full
curves), and 50 ms and $\alpha = 2$ (dot-dot-dot-dashed curves) ($\beta =1$
in the above cases), and for $\langle p\rangle = 10$ ms, $\alpha = 2$, and
$\beta = 0.1$ (thick dashed curve).}    
\label{fig2}
   \end{figure}

At the end of the optically thick phase and during the optically thin phase,
the pulsar-supernova envelopes should become also the sources of 
$\gamma$-rays and neutrons. Production of these neutral particles has been
already considered in a similar supernova model (with the difference that
particles are accelerated in the inner pulsar magnetosphere, Protheroe et
al.~1998, Bednarek \& Protheroe~1997). The contribution of these neutral
particles to the $\gamma$-ray background and cosmic rays 
in terms of the present model will be discussed in a future paper. 

We have discussed also the problem whether the estimated extragalactic neutrino
background from the whole population of pulsars in the universe can be
detected by the planned experiments. The full horizontal line in Figs.~ 1 and
2 shows the detection limit of the  IceCube detector (Spiering~2000). It looks
that ENB from pulsars should be easily observed by the IceCube if significant
part of pulsars are born with initial periods shorter than $\sim 10$ ms in the
case of uniform distribution (see Fig.~1), 
and if the mean pulsar period is close to 10 ms in the case of the gamma
distribution (see Fig.~2), assuming that $\eta\sim 1$.  Also the planned AUGER
experiment with the  detection limits estimated by Bertou et al.~2001
may observe ENB from pulsars if
the significant amount of pulsars is born with the initial periods of the
order of a few ms as postulated by the recent models for the pulsar origin of
the highest energy cosmic rays (BEO, De Goubeia Dal
Pino \& Lazarian~2000). However, such detection would be possible if
the hypothesis of $\nu_\mu\rightarrow \nu_\tau$ oscillations with full mixing 
is correct. In such case the expected flux of tau neutrinos, compared to
the calculated flux of muon neutrinos, is above the sensitivity limit of the
AUGER detector for the case of pulsars born with the milisecond periods (see
lower dotted curve in Fig.~1, taken from Bertou et al.~2001).

\begin{acknowledgements}
I thank the referee for comments and suggestions. This work is supported by 
the Polish KBN grant No. 5P03D02521. 
\end{acknowledgements}


\begin{thebibliography}{}

\bibitem{a98} Andersson, N. 1998, ApJ, 502, 708

\bibitem[2001]{bb01} Beall, J.H., Bednarek, W. 2001, ApJ, submitted (BB)

\bibitem[2001]{bp97} Bednarek, W., Protheroe, R.J. 1997, PRL, 79, 2616

\bibitem[2001]{bp01} Bednarek, W., Protheroe, R.J. 2001, Astropart.Phys., in
press

\bibitem[1978]{bp78} Berezinsky, V.S., Prilutsky, O.F. 1978, A\&A, 66, 325

\bibitem[2001]{betal01} Bertou, X., Billoir, P., Deligny, O., Lachaud, C.,
Letessier-Selvon, A. 2001, Astropart.Phys., in press

\bibitem[2000]{beo00} Blasi, P., Epstein, R.I., Olinto, A.V. 2000, ApJ, 533,
123 (BEO)

\bibitem{bl86} Burrows, A., Lattimer, J.M. 1986, ApJ, 307, 178

\bibitem[2000]{dl00} De Goubeia Dal Pino, E.M., Lazarian, A. 2000, ApJ, 536,
L31

\bibitem[1981]{ft81} Fichtel, C.E., Trombka, J.I. 1981, Gamma Ray
Astrophysics, (NASA SP-453, Washington, DC), 192

\bibitem[2001]{gl01} Giller, M., Lipski M. 2001, Proc. 28th ICRC (Hamburg),
p.~2092

\bibitem[2000]{ktn00} Kobayashi, C., Tsujimoto, T., Nomoto, K. 2000, ApJ,
539, 26 

\bibitem[2000]{lm00} Learned, J.G., Mannheim, K. 2000, Ann.Rev.Nucl.Part.Sci.,
50, 679

\bibitem{lom98} Lindblom, L., Owen, B.J., Morsink, S.M. 1998, PRL, 80, 4843

\bibitem[2001]{mpr01} Mannheim, K., Protheroe, R.J., Rachen, J.P. 2001,
Phys.Rev., D63, 23003

\bibitem[1987]{n87} Narayan, R. 1987, ApJ, 319, 162

\bibitem[1998]{p98} Protheroe, R.J. 1998, astro-ph/9809144

\bibitem[1998]{pbl98} Protheroe, R.J., Bednarek, W., Luo, Q. 1998, Astropart.
Phys., 9, 1

\bibitem[2000]{s00} Spiering, C. 2000, Nucl.Phys.Suppl., 91, 445

\bibitem{wb99} Waxman, E., Bahcall, J. 1999, Phys.Rev. D, 59, 023002

\bibitem{wyhw00} Wheeler, J.C., Yi, I., H\"oflich, P., Wang L. 2000, ApJ, 537,
810

\bibitem{wlw93} Woolsey, S.E., Langer, N., Weaver, T.A. 1993, ApJ, 411, 823

\end{thebibliography}
\end{document}